\RequirePackage{fix-cm}
\documentclass[natbib,twocolumn]{svjour3}          
\smartqed  
\usepackage{graphicx}
 \journalname{Space Science Reviews}
\newcommand{\lum}{erg s$^{-1}$}

\def\ltsima{$\; \buildrel < \over \sim \;$}
\def\lsim{\lower.5ex\hbox{\ltsima}}
\def\gtsima{$\; \buildrel > \over \sim \;$}
\def\gsim{\lower.5ex\hbox{\gtsima}}

\def\msun{~M_{\odot}}

\def\mdot {\dot M}

\begin{document}

\sloppypar
\title{Magnetic fields of neutron stars in X-ray binaries
}


\author{Mikhail Revnivtsev        \and
        Sandro Mereghetti 
}


\institute{M. Revnivtsev \at
              Space Research Institute, Russian Academy of Sciences, \\
              Profsoyuznaya 84/32, Moscow, Russia, 117997 \\
              \email{revnivtsev@iki.rssi.ru}           
           \and
           S. Mereghetti \at
INAF, IASF-Milano, v. Bassini 15, I-20133 Milano, Italy\\
              \email{sandro@iasf-milano.inaf.it}  
              }

\date{Received: date / Accepted: date}

\maketitle

\begin{abstract}
A substantial fraction of the known neutron stars resides in  X-ray binaries --  systems in which one  compact object accretes matter from a companion star.   Neutron stars in  X-ray binaries have magnetic fields among the highest found in the Universe, spanning at least the range from  $\sim10^8$  to several 10$^{13}$ G.  The magnetospheres around these neutron stars have a strong influence on the accretion process, which powers most of their emission.  The magnetic field intensity and geometry, are among the main factors responsible for the large variety of spectral and timing properties observed in the X-ray energy range, making these objects unique laboratories to study the matter behavior and the radiation processes in magnetic fields unaccessible on Earth.  In this paper we review the main observational aspects related to the presence of magnetic fields in neutron star  X-ray binaries and some methods that are used to estimate their strength.
\end{abstract}

\section{Introduction}
X-ray binaries (XRBs) were the first astrophysical objects in which relativistic compact stars (neutron stars (NSs) and black holes) were detected. In 1962 the pioneering observations of the sky in the X-ray energy range  led to the discovery  of the bright source Sco X-1 \citep{giacconi62}, which is now known to contain an accreting NS. However,  at that time the origin of the X-ray emission and the nature of the Sco X-1 source were completely unknown. Thus, the first ones to recognize the existence of NSs were radio astronomers,  who  discovered the radio pulsars  a few years later \citep{hewish68,gold68}. Measurements of the  spin-down of radio pulsars, interpreted as a loss of rotational energy due to the emission of magneto-dipole radiation, allowed the first estimates of their magnetic fields \cite[e.g.][]{gunn69}. These  were in agreement with earlier predictions based on the 
assumption that  the magnetic flux of stars collapsing into NS is conserved,  giving rise to NS magnetic fields as high as $B\sim 10^{12}-10^{16}$ G \citep{ginzburg64,woltjer64}. 

The observations obtained by UHURU, the first satellite for X-ray astronomy,  showed that many XRBs contain NSs with magnetic fields powerful enough to overcome the enormous gravitational drag of the NS and disrupt the flow of accreting matter, thus  creating anisotropic emission patterns. These allowed rotating accreting NSs to be detected as regularly pulsating X-ray sources through a lighthouse effect \citep{giacconi71,schreier72,tananbaum72,lamb73,davidson73}.

Since the time of these early discoveries,  significant advances in observational astrophysics have provided an extremely rich dataset on NSs. Differences in their intrinsic properties (ages, magnetic fields, spin-periods, etc.), as well as in their environment (e.g., isolated or in binaries, density of interstellar medium, location in globular clusters, etc.), give rise to a variety of manifestations that can be studied in the whole electromagnetic spectrum.  
In this paper we concentrate on the NSs whose emission is mainly powered by accretion of matter provided by a companion star. The accreting matter acts as a probe in the regions of our interest, in particular around magnetic NSs.  Observations in  X-rays (from $\sim$1 keV to a few hundred keV) are one of the most informative ways to explore the properties of NSs
because this is the energy range  where the  NS surface  and the matter in its immediate vicinity emit a significant part of their bolometric luminosity. 

In  Section 2 we introduce a few basic equations relevant for the physics of accretion and give a brief overview of the main properties of XRBs containing NSs.   
In the following Sections we focus on a few specific aspects closely related to the presence of magnetic fields: the observation of cyclotron lines (Sect. \ref{sec:cycl}), and the interaction between the accretion flow and the NS magnetosphere (Sect. \ref{sec:interaction},\ref{sec:aperiodic}, and \ref{sec:propeller}).

\section{Accretion-powered  neutron star X-ray binaries}
\label{sec:main_prop}

X-ray binaries  are powered by accretion onto a compact object which gravitationally captures part of the mass lost by its companion star. The early evidence that  many of the observed properties of XRBs   depend on the type of companion star led to their classification into the two main classes of high mass X-ray binaries (HMXBs) and  low mass X-ray binaries (LMXBs), based on the mass of the companion star.
Extensive reviews of this vast subject can be found,  e.g., in
\cite{lewin_vdk10}.  
Progress in observations led to the discovery of a large number of XRBs in our Galaxy and in the Magellanic Clouds \cite[see, e.g., ][for recent catalogs of XRBs]{liu06,liu07}. At the moment, the most sensitive surveys of the whole Galactic plane obtained in hard X-rays have detected about two hundred  sources, nearly equally divided between HMXBs and LMXBs \citep{krivonos10,bird10,baumgartner13}.

Before giving an overview of the main properties of XRBs containing NSs, we remind a few basic concepts and definitions that will be used in the following sections   \cite[see, e.g.,][for a more extensive discussion]{frank92}.
We denote with M and R the NS mass and radius, and assume in all the numerical formulae the values  $M=1.4 M_\odot$  and  $R=12$ km. The NS moment of inertia is   $I\approx 0.4M R^2$, and its angular velocity is  $\omega=2\pi\nu=2\pi/P$.  
The magnetic dipole moment is  $\mu = B R^3/2$,  where $B$ is the field strength at the NS magnetic poles.

The  accreting matter, on its way toward the compact object,  releases a large amount of  gravitational energy, heats up and emits mostly in the X-ray energy range.  In the case of  an accreting NS, the accretion luminosity is given by:
\begin{equation}
L_{\rm X}\approx  {G M  \dot{M}\over{R}} \approx 1.5\times10^{20} ~\dot{M} ~~\rm{erg~s^{-1}},
\label{acc}
\end{equation}

\noindent
where $\dot{M}$ is the mass accretion rate in g s$^{-1}$. This corresponds to a $\sim$17\% efficiency of conversion of the rest mass energy of the accreting matter.

An upper limit to the accretion luminosity is given by the Eddington luminosity, $L_{\rm Edd}$, at which the radiation force equals the gravitational attraction, thus stopping the accretion flow:
\begin{equation}
L_{\rm Edd}=   {4 \pi G M  m_{\rm p} c \over {\sigma_{\rm T}}} \approx 1.3 \times10^{38} {M \over{\msun}} ~~\rm{erg~s^{-1}},
\label{edd}
\end{equation}

\noindent
($c$ is the velocity of light, $m_{\rm p}$ is the proton mass, and $\sigma_{\rm T}$ is the Thompson cross section). The typical luminosities observed in bright XRBs are of the order of $\sim10^{36}-10^{38}$ erg s$^{-1}$.  They  imply  accretion rates of   $\sim10^{16}-10^{18}$g s$^{-1}$ (corresponding to $\sim 10^{-10}-10^{-8}$  $\msun$ yr$^{-1}$). Much lower luminosities,  down to $\sim10^{31}$ erg s$^{-1}$,  are seen in the quiescent states of transient sources (note that other processes besides accretion might be at work to power  these low luminosities).

To discuss the effects produced by the NS magnetic fields on the flow of accreting matter, hence on the X-ray emission properties, it is useful to consider a few characteristic radii around the compact object.  Their relative positions,  which  depend mainly on $\omega$, $\mu$,  and density of the inflowing mass, determine the appearance of the NS \citep[see, e.g.,][]{lipunov92,campana98review,bozzo08}.

The accretion radius, $r_{\rm a}$,  defines the region in which the matter is gravitationally captured by the compact object.  In the case of a NS accreting from a stellar wind with velocity $v_{\rm w}$, it is given by:

\begin{equation}
r_{\rm a} = 2GM/(v_{\rm w}^2 + v_{\rm orb}^2),
\label{ra}
\end{equation}

\noindent
where    $v_{\rm orb}$ is the orbital velocity of the NS (usually negligible compared to the wind velocity $v_{\rm w}$).

The magnetic  energy density, $B^2/8\pi$, around a magnetized  NS rapidly rises towards the compact object (as $\propto r^{-6}$ in the case of a dipole  and even faster in the case of higher multipole fields). A  simple estimate of the distance from the NS where the magnetic pressure becomes equal to the ram pressure of the accretion flow gives the magnetospheric radius \citep[see, e.g.,][]{pringle72,lamb73,kluzniak07}:

\begin{equation}
r_{\rm m}= \xi \left( {\mu^4\over{G M\dot{M}^2}} \right)^{1/7},
\label{rm}
\end{equation}

\noindent
where  a dipolar magnetic field  has been assumed. The  parameter $\xi$ depends on the configuration of the flow:  it is  $\sim$1 for a spherical inflow  \citep{pringle72},  while it can be smaller, $\xi\sim$0.4, in the case of a Keplerian disk  \citep{ghosh79}. 
Within $r_{\rm m}$ the motion of the matter is strongly influenced by the geometry of the magnetic field, which rotates rigidly with the star.  

Two other relevant radii are the  corotation radius, $r_{\rm co}=(GM/\omega^2)^{1/3}$,  and the  light-cylinder radius, $r_{\rm lc}=c/\omega$, 
where the linear velocity of the rigidly corotating magnetosphere  equals, respectively,  the Keplerian velocity and the speed of light.

\subsection{High Mass X-ray Binaries}

HMXBs are systems in which the mass donor is typically an OB star with mass greater than 6--7 $M_\odot$.
Known HMXBs have orbital periods from 0.2 days (Cyg X-3, with a Wolf-Rayet companion) to almost one year ($P_{\rm orb}\sim 262$ days of SAX J2239+6116) and spin periods ranging  from 0.7 s (SMC X-1) up to $\sim$14 ks (1H 1249-637).
The detection of X-ray pulsations in most of these systems was one of the first indications for the presence of magnetized NSs, with fields  sufficiently strong ($\sim10^{11-13}$ G) to channel the accretion flow onto the magnetic poles  and produce beamed X-ray emission.  These magnetic fields are of the same order of those found in young radio pulsars, consistent with the evidence that HMXBs have ages smaller than few tens of Myrs, as implied by the presence of massive stars in these systems.

\begin{figure*}
\includegraphics[width=\textwidth,bb=165 244 673 349,clip]{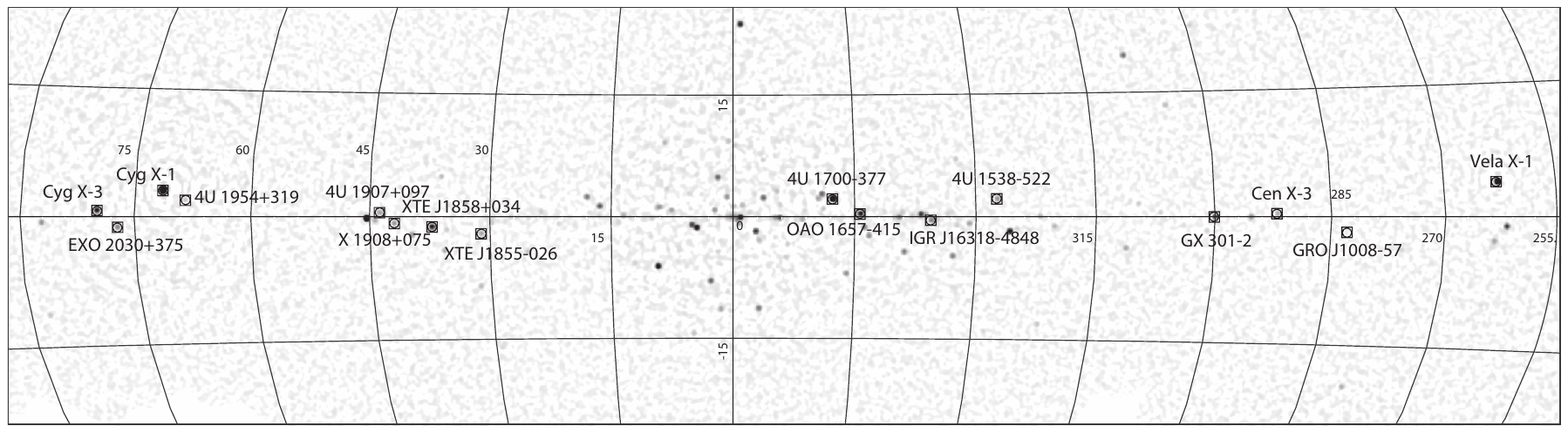}
\caption{Image of the central region of the Galactic plane (in Galactic coordinates), obtained by the INTEGRAL observatory in the energy band 17-60 keV. The  positions of some of the brightest HMXBs in our Galaxy are indicated. The vast majority of them contain accreting magnetic NSs.}
\label{map}
\end{figure*}

Extensive surveys of the Galaxy  \cite[e.g.][]{lutovinov13} have revealed several tens of persistently bright HMXBs with luminosity  $L_{\rm X}\sim10^{36}-10^{37}$ erg s$^{-1}$ (positions of some brightest sources on the sky are shown in Fig. \ref{map}). Many of them are located in the spiral arms, associated to regions of recent star formation.
Persistent HMXBs typically have  OB supergiant companion stars, characterized by strong stellar winds with velocity $v_{\rm w}$ of the order of a thousand km s$^{-1}$ and mass loss rates $\mdot_{\rm w}\sim10^{-7}-10^{-5}$ $\msun$ yr$^{-1}$. Such winds provide the matter which is accreted by the NS. Several HMXBs with OB companions are strongly absorbed in the soft X-ray energy range, due to the presence of dense circumstellar material, and could be discovered only recently thanks to observations in the hard X-ray range \citep{rev03,walter03,kuu05}.

SFXTs (Supergiant Fast X-ray Transients) have OB-type supergiant companions, like those of the persistent HMXBs, but they are characterized by strong X-ray outbursts of short duration spanning a very large dynamic range -- up to 3 or 5 orders of magnitude from quiescence to the peak of their outbursts \cite[see, e.g.,][]{sguera06,sidoli2012}. SFXTs  spend most of the time at very low luminosities, and sporadically emit outbursts with duration of the order of few hours or even less \citep[see e.g. reveiw in][]{paizis14},  although sometimes superimposed on longer periods of activity.  Due to the rarity and short duration of their outbursts,  SFXTs escaped detections until recently. Thanks to the extensive coverage of the Galactic plane with the INTEGRAL observatory, this class has now grown up to a dozen of sources \citep{romano14}. The mechanism at the basis of SFXT outbursts is not yet understood and several models have been proposed \cite[e.g.][]{zand05,sidoli07,walter07,grebenev07,ducci09}.


The largest population of HMXBs consists of NSs with Be type companions \cite[e.g.][]{reig11}. Most of them are transient systems, which spend a large fraction of time in a low luminosity (or off-) state. Their outbursts are related to the presence of the dense equatorial outflows in the winds of Be stars, which are responsible for the H$_{\alpha}$ emission lines seen in the optical spectra of these stars.  The transient behavior of these XRBs occurs because the NS acquires a higher accretion rate when it crosses the Be equatorial disk and/or because the disk ejection itself is subject to long term variability.

Most NSs in  HMXBs accrete matter gravitationally captured from the stellar winds of  massive companions which  underfill their Roche-lobe \cite[e.g.][]{davidson73}.
More rarely, the accretion flow is significantly modified by a Roche-lobe influence \cite[e.g.][]{lamers76}. 

The strong fields of NSs channel the accreting matter onto the magnetic poles, forming an accretion column \citep[e.g.][]{basko76,lyubarskii88,mushtukov14}. Most of the kinetic energy of the matter falling onto the NS surface should be released as radiation of the accretion column, with a luminosity of the order of $\sim 0.2 \dot{M}c^2$ (eq. (\ref{acc})). The free fall velocity near the NS surface is $\sim 0.6c$, which corresponds to effective temperatures larger than $10^{12}~K$ ($>100$ MeV). The observed emission, however, is more concentrated in the 1-20 keV energy band  and the spectrum is significantly different from that of a simple blackbody emission -- the spectrum is typically a power law ($dN/dE\propto E^{-\Gamma}$) with photon index $\Gamma\sim1$  up to energies 10-30 keV and an exponential cutoff at higher energies \cite[see, e.g., ][]{white83,nagase89,filippova05,caballero12}. 
The emission pattern is strongly anisotropic, giving rise to easily detectable X-ray pulsations due to the rotation of the NSs.  In most sources the pulse profiles show significant variations as a function of energy, luminosity level, orbital phase, and time. 
 
The details of the spectral formation in HMXBs are not yet fully understood, in spite of very serious efforts in this direction \citep{nagel81,lyubarsky82,meszaros85,burnard91,shibanov92,klein96,becker05,becker07}. Modelling the spectra of the accretion column on NS surfaces of X-ray pulsars is complicated by: a) the presence of powerful radiation pressure, which determines the dynamics of the settling flow, b) the strong magnetic field,
which modifies all scattering cross sections and thus the radiation transfer, and c) the presence of fast bulk motion of the infalling matter, which contributes to the Compton upscattering of the outgoing radiation. All these complications should be solved self-consistently.

\subsection{Low Mass X-ray Binaries}

The companions of NSs in LMXBs are typically  late type dwarf stars with mass below 1--2 $\msun$. Most LMXBs have orbital periods shorter than one day. The companion stars fill their Roche-lobe and accretion proceeds with the formation of a disk around the compact object.
LMXB can have ages up to a few  Gyrs, as reflected by their association with the old stellar population in the Galactic bulge and by their presence in globular clusters.
The existence of dynamically important magnetic fields in LMXBs, able to channel the accretion flow onto the NS magnetic poles, has been under discussion for a long time because regular pulsations were originally seen only in a few peculiar members of this class, like Her X-1 ($P$=1.24 s), 
4U 1626--67 ($P$=7.7 s) and  
GX 1+4 ($P$=130 s)  \cite[see][and ref. therein]{liu07}.
Extensive searches for pulsations in other bright LMXBs were unsuccessful \cite[see, e.g.,][]{mereghetti87,vaughan94} and the presence of NSs, rather than accreting black holes, in these systems could only be established through the observation of type I X-ray bursts - thermonuclear explosions in the accreted matter which accumulates onto the NS surface \cite[see, e.g., reviews in][]{lewin93,cumming04}.  

Thanks to the large effective area provided by the instruments on the RXTE satellite,  X-ray pulsations were finally discovered in LMXBs \citep{wijnands98}.  At the moment we know 15 systems of this class in which the NSs create persistent or intermittent X-ray pulsations with periods of a  few milliseconds  \cite[see, e.g., ][]{patruno12} and several sources which show millisecond pulsations  during type I bursts \cite[e.g.][]{strohmayer01}. These findings show that the NSs in LMXBs have spin periods much shorter than those in HMXBs. Their fast rotation is due to the spin-up torque applied on the NS  by the rapidly rotating Keplerian accretion disk, which extends close to the star surface. This implies magnetic fields B$\sim10^{8-10}$ G,  smaller than those found in HMXBs, which could be a consequence of the old age of the NSs in LMXBs, possibly coupled to the field decay induced by the accretion process \cite[e.g.][]{taam86,harding06}.  Such  low magnetic fields are at the basis of the well established recycling scenario, which explains the millisecond radio pulsars as old NSs spun-up by the accretion process in LMXBs \citep{bk76,bhattacharya91,srinivasan10}.

The weak magnetic field of these NSs is also at the basis of the different X-ray spectral properties of LMXBs and HMXBs. In general, at energies below $\sim$20 keV the spectra of LMXBs are much softer than those of HMXBs. This is due to the fact that a large fraction of the observed  X-rays  originates in optically thick accretion disks,  which in LMXBs extend close the compact object where they reach temperatures of only a few keV.

\section{Resonant cyclotron features in the X-ray spectra of  accreting NSs}
\label{sec:cycl}

The  energy spectra of several accreting magnetic NSs show the presence of absorption features at energies of tens of keV. Discovered more than thirty years ago \citep{trumper78,wheaton79}, they were already expected by theoreticians \citep{gnedin74}. These features  were immediately interpreted as due to cyclotron resonant scattering of the outgoing radiation by electrons in a strong magnetic field  and provided the first direct evidence for magnetic fields of the order of 10$^{12-13}$ G in the line-forming regions. The  line energies are determined by the separation of the quantized Landau energy levels of electrons, which, in a magnetic field of strength $B$, is given by:

\begin{equation}
E_{\rm c}=\hbar {eB\over{m_e c}} =11.6 {B\over{10^{12}\textrm{G}}}   ~\textrm{keV}
\label{cyc}
\end{equation}

\noindent
Due to the gravitational redshift $z$ at the position where the scattering occurs, the lines in the  observer's frame of reference appear at  $E_{\rm c}/(1+z)$. 

\begin{table*}[htb]
\caption{XRBs with cyclotron resonance features (updated from \citealt{caballero12}). The 5$^{th}$ coumn gives the number of harmonics. The + and --  symbols in the 6$^{th}$ column indicate a positive or negative correlation between  $E_{\rm c}$ and $L_{\rm X}$ (tentative reporst are indicated by a question mark);  the = symbol indicates that  $E_{\rm c}$ has been seen to stay constant during  luminosity changes. 
}
\begin{center}
\begin{tabular}{lclccll}
\hline
Name & Type & $E_{\rm cyc}$ & B/(1+z) &  Num. &   $E_{\rm c}$--$L_{\rm X}$   & References \\
         &          &   keV           & $10^{12}$G&  harm.    & correl.      &             \\
\hline
Swift J1626.6--5156 & Be &10    &0.9  &1& + & \cite{decesar13}\\
4U 0115+63            & Be &11    &  1  &4& --? & \cite{wheaton79,muller13}\\
KS 1947+300            & Be &12 &1     &  &    & \cite{fuerst14}\\
IGR J17544--2619   &SFXT& 17 & 1.5  &  &    &\cite{bhalerao14}\\
4U 1907+09             & SG &18  &1.5    &1&    &\cite{makishima92,cusumano98}\\
4U 1538--52            & SG &23  &2    &  & +  & \cite{clark90,hemphill14}\\
Vela X-1                   & SG &25  &2.1    &1&+ &  \cite{kendziorra92,fuerst14a}\\
V 0332+53               & Be &26  &2.2    &2&--&  \cite{makishima90,tsygankov10}\\
X Per                       & Be & 29? & 2.5   &  &  &\cite{coburn01,doroshenko2012b}\\
Cen X-3                  & SG  & 30 &  2.6  &   &  &   \cite{santangelo98,suchy08}\\
Cep X-4                   & Be  &31 &2.7  &  &  &\cite{mihara91,mcbride07}\\
RX J0520.5--6932  & Be & 32 & 2.8  & &  &  \cite{tendulkar14} \\
RX J0440.9+4431   & Be &32    & 2.8   &  &  & \cite{tsygankov12}\\
IGR J16493--4348  & SG & 32   &   2.8      &  &  & \cite{dai11} \\
MXB 0656--072     & Be & 33    & 2.8  & &   &\cite{heindl03,mcbride06}\\
GX 301--2            & SG & 35     &3 & &   & \cite{makishima92,krey04}\\
XTE  J1946+274     & Be & 36    &3.1   &  &+? & \cite{heindl01,muller12}\\
4U 1626--67        & LMXB& 37  &3.2  & &=&  \cite{orlandini98,camero2012}\\
Her X-1               & LMXB & 41  & 3.5 &  &+&\cite{trumper78,staubert07}\\
A 0535+26          & Be & 46     &4   &1&=&\cite{kendziorra92,muller13b}\\
GX 304--1          & Be & 51      &4.4  &   &+& \cite{yamamoto11,klochkov12}\\
1A 1118--61      & Be & 55      &4.7&  & &\cite{doroshenko10,maitra12}\\
GRO J1008--57  & Be &  78   & 6.7 &  & &\cite{shrader99,bellm14}\\
\hline
\end{tabular}
\end{center}
\end{table*}

\begin{figure}[htb]
\includegraphics[width=\columnwidth,bb=141 89 700 520,clip]{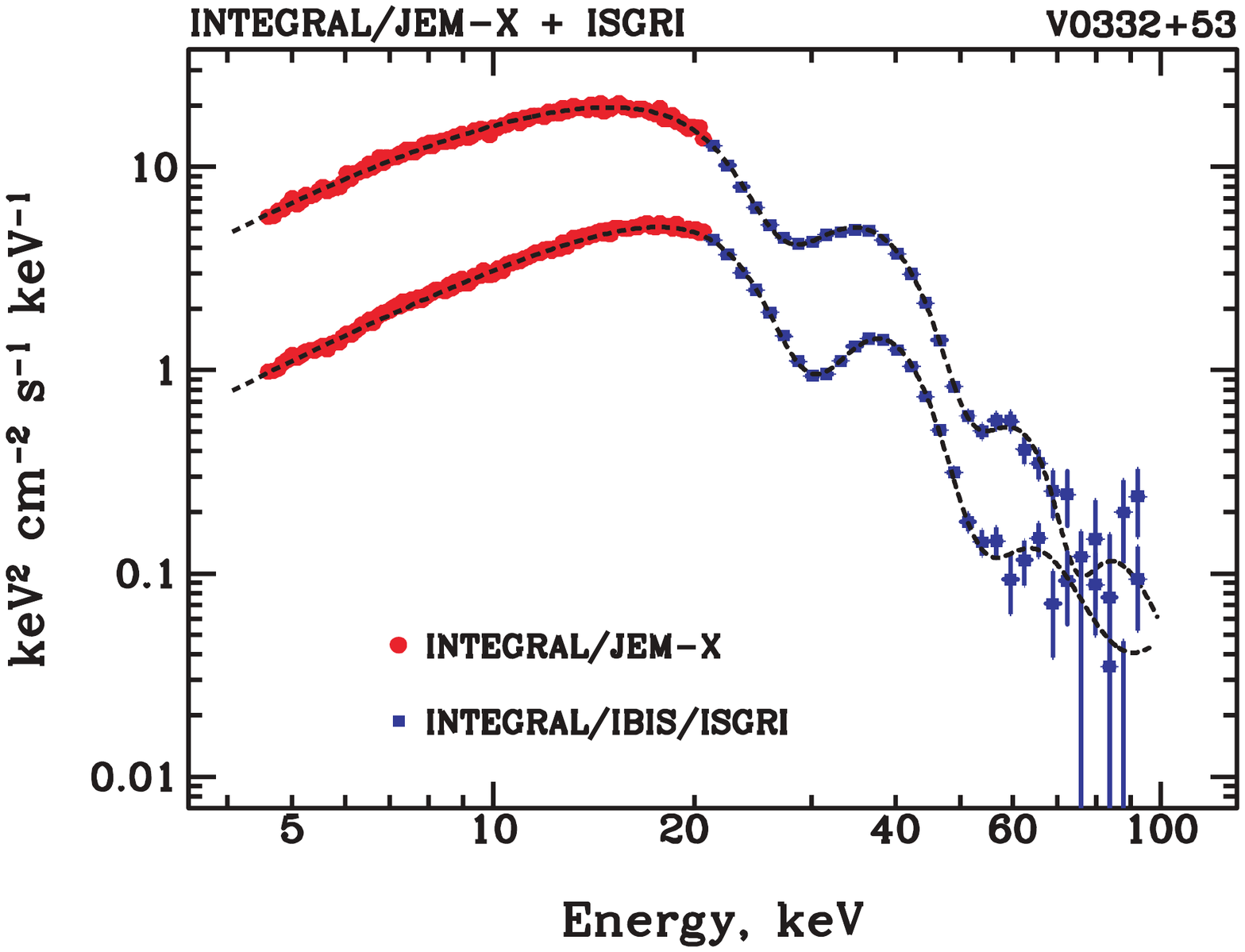}
\caption{Energy spectrum of the accretion powered pulsar V0332+53 at two intensity levels measured with the INTEGRAL observatory.  At least two cyclotron absorption lines are clearly visible  (from \citealt{tsygankov06}).}
\label{cycl_lines}
\end{figure}

Over the last decades, cyclotron absorption features have been detected from more than  twenty sources (see Table 1). 
In a few sources one or more harmonics of the fundamental line are also seen \citep{santangelo99,tsygankov06}. One of the best examples is presented in Fig.\ref{cycl_lines}.  Note that the line energies given in  Table 1 are only approximate values, since many sources show line energy variations as a function of the spin phase, time, and/or flux. In addition, the exact values of the line parameters derived from the spectral fits depend on the spectral model used to describe the continuum.  
Despite these uncertainties, it is clear that the observed line energies imply magnetic fields in the range $\sim10^{12}-10^{13}$ G, both for NSs with Be and supergiant (SG) companions. Note that Her X-1 and GX 1+4, the only LMXB showing cyclotron lines,  are not representative members of the LMXB population: they  resulted from different evolutionary histories and have magnetic fields similar to those of HMXBs. Among the most recent discoveries, it is worth noting that the first cyclotron line discovered in a member of the SFXT class indicates a ``standard'' field value   \citep{bhalerao14}. This  does not support the models invoking strong magnetic fields to explain these sources \cite[e.g.][]{bozzo08}.

Variations of the line properties as a function of the source luminosity are particularly interesting since, in principle, they offer the possibility to study how the geometry of the accretion region changes at different mass accretion rates. 
In the simplest interpretation, the observed variations of line energy, $E_{\rm c}$, as a function of the luminosity,  $L_{\rm X}$, trace the height of the X-ray emitting shock in the accretion column.  At luminosities above the local critical value ($L_{\rm X} \gsim 10^{37}$  erg s$^{-1}$), as the mass accretion rate increases the radiation pressure moves the shock farther from the NS surface, where the magnetic field is weaker. Hence an anticorrelation between $E_{\rm c}$ and $L_{\rm X}$ is expected \citep{basko76,burnard91}.  This behavior has been observed in V0332+53 \citep{tsygankov06} and in 4U 0115+63 \citep{nakajima06}.  However, the analysis of another outburst from the latter source indicates that the reported anticorrelation might be an artifact caused by an improper modelling of the underlying variations of the continuum spectrum \citep{muller13}. 
A positive correlation between $E_{\rm c}$ and $L_{\rm X}$ has instead been seen in Her X-1 \citep{staubert07} and in a few other sources (see Table 1). Their different behavior has been attributed to the lower luminosity of these sources, in which gas pressure and Coulomb scattering become important in shaping the geometry and emission pattern of the accretion column \citep{becker12}.

There is general agreement that the observed lines are due to resonant electron scattering in a strong magnetic field and models based on
the transmission of radiation through a highly magnetized medium in the accretion column   can account for  most of the observations \cite[see e.g.][]{schonherr07,nishimura08}. However, the simple picture outlined above predicts a luminosity-dependence of  $E_{\rm c}$   at high accretion rates stronger than the observed one (typically $E_{\rm c}$ varies by less than 20-30\% when the luminosity changes by an order of magnitude). A recent model to overcome these problems is presented by \citet{nishimura14}, while a  different explanation is based on  the idea that the cyclotron lines are formed as a result of reflection of the accretion column X-ray emission from magnetized matter on the NS surface \citep{poutanen13}.

\section{Interaction between NS magnetosphere and accretion flow}
\label{sec:interaction}

\begin{figure*}[htb]
\includegraphics[width=\columnwidth,bb=129 197 566 550,clip]{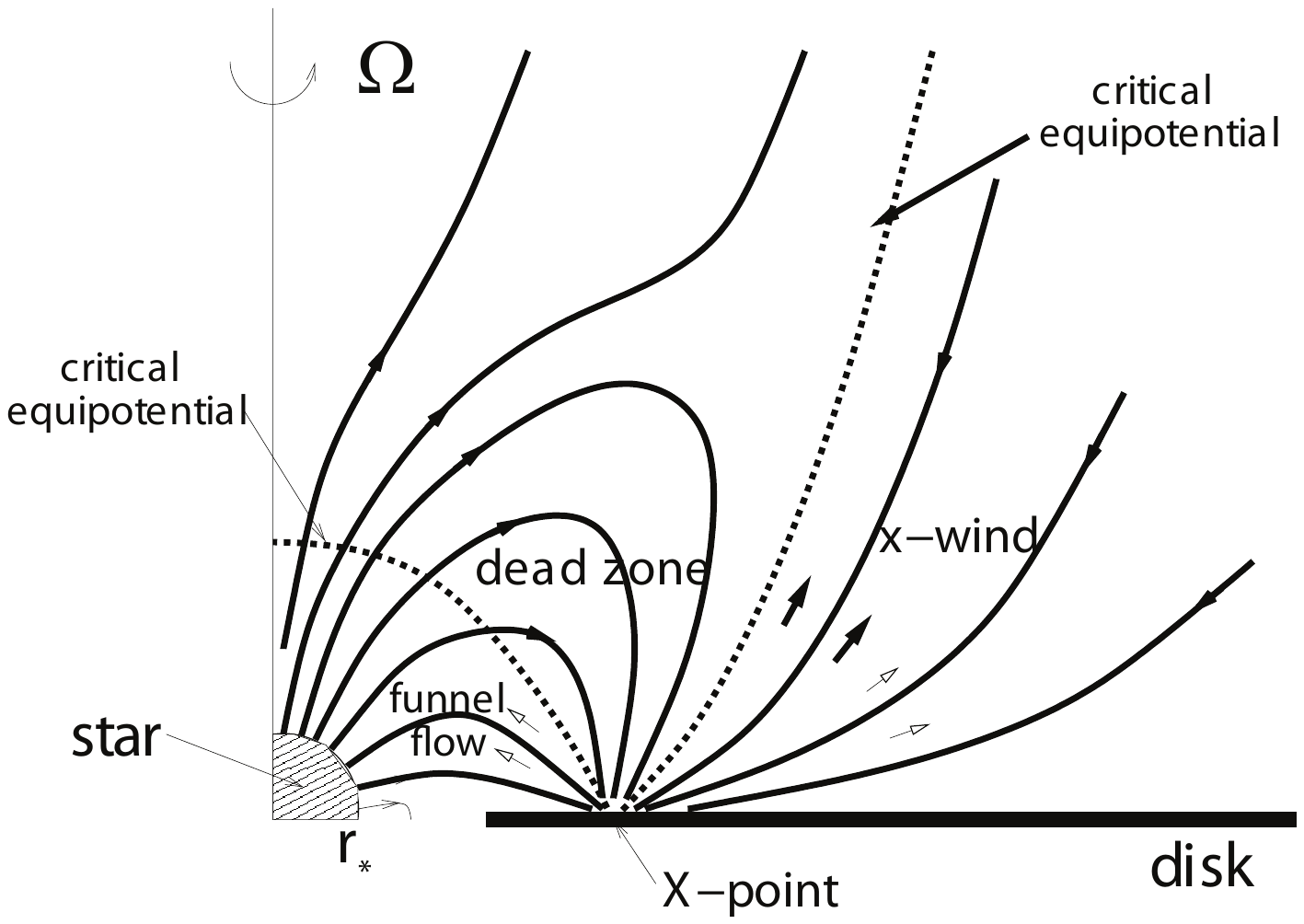}
\includegraphics[width=\columnwidth,bb=224 70 750 550,clip]{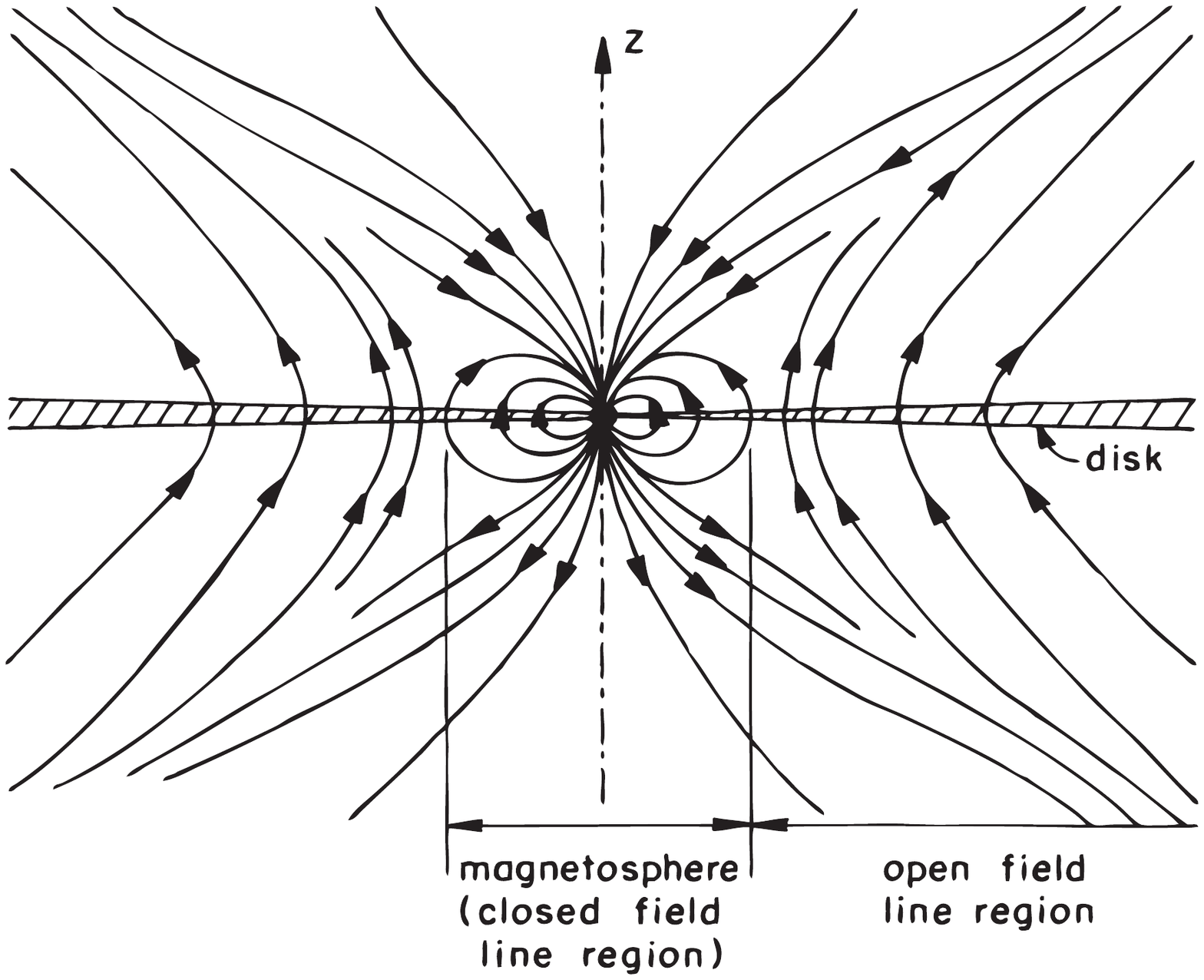}
\caption{Models of magnetosphere-accretion disk interaction of \cite{shu94} (left panel) and \cite{lovelace95} (right panel). Both models imply that the accretion disk matter is frozen into the NS magnetic field lines only over a relatively small transition region.}
\label{schemes}
\end{figure*}

Although the details of the interaction between the rotating magnetosphere and the accretion flow are quite complicated and not yet fully understood \cite[see, e.g.,][]{uzdensky04,lai14}, it is generally assumed that  accretion onto the NS surface can occur if the magnetospheric radius (eq.(\ref{rm})) is smaller than the corotation radius. 
If, instead, $r_{\rm m}>r_{\rm co}$, the centrifugal barrier should stop the flow and inhibit continuous accretion \citep{illarionov75,stella86}.  
In this case, non-stationary accretion might still proceed as the matter piles up around the intrinsically unstable magnetospheric boundary. Depending on its specific angular momentum,  the captured plasma may form a quasi-spherical shell \citep{pringle72,ikhsanov07,shakura12} or reside in a so-called ``dead accretion disk" \citep{ss77,baan79,dangelo10,dangelo12}.

Accretion   can proceed in two ways: either quasi-spherically or via a geometrically-thin disk.
The latter scenario typically occurs when the mass donor star fills its Roche-lobe, as it happens in most LMXBs.   
In the case of wind-accreting HMXBs,  quasi-spherical accretion can occur if  the angular momentum
of the mass captured within the accretion radius is sufficiently small. This can be evaluated by considering the circularization radius, i.e. the
minimum distance from the NS which  the accreting matter can reach without loss of angular momentum. 
The circularization radius, $r_{\rm circ}$, can be roughly estimated by equating the specific angular momentum of the captured stellar wind matter to that on a Keplerian orbit (see, e.g., \citealt{frank92}):
\begin{equation}
j\sim {\pi r_{\rm a}^2\over{2 P_{\rm orb}}} \sim \sqrt{GM r_{\rm circ}},
\end{equation}

\noindent
This leads to:

\begin{equation}
r_{\rm circ}\sim {4 \pi^2 G^3 M^3\over{P_{\rm orb}^2 v_{\rm w}^8}} 
\label{rcirc}
\end{equation}

\begin{equation}
r_{\rm circ} \sim  3.3\times10^{6}   \left({1~~{\rm day}\over{ P_{\rm orb}}}\right)^2  \left({1000~~{\rm km~s}^{-1}\over{v_{\rm w}}}\right)^8 ~\textrm{cm}
\label{rcirc2}
\end{equation}

\noindent
For typical values of NS magnetic fields, mass accretion rates, and  stellar wind velocities of young OB-type stars,
it turns out that, for wide binaries ($P_{\rm orb}>$ a few days),  a disk cannot form because $r_{\rm circ}$ $<$ $r_{\rm m}$.  In this case  the accreting matter should  settle onto the rotating magnetosphere in some kind of a quasi-spherical flow. 

The resulting interaction is a long standing problem and  still  a matter of debate \cite[see, e.g.,][]{arons76,lamb77,davies79,illarionov90,ziolkowski85,bozzo08,shakura12,shakura13}. 
Unsettled questions on details of the magnetosphere-flow interaction sometimes lead to controversial suggestions about the presence of utrastrong magnetic fields in pulsars with long spin periods  \cite[see e.g.][]{finger10,doroshenko10b,reig12,ho14}. The root of this problem is in the assumptions on the values of the torques exerted by the accreting matter settling to the NS magnetosphere. 

Contrary to the disk accretion case,  where the torques are relatively well understood and the predicted scalings confirmed by observations (see Sect. \ref{sec:interaction}),
the issue of the net torque acting on the NS is much less clear in the case of quasi-spherical accretion  (see discussion of this problem in \citealt{shakura12,shakura13,postnov14}).

In the cases of accretion from a companion star in tight binaries (where Roche-lobe overflow occurs), from the equatorial disks around Be stars, and from low mass companions, the specific angular momentum of the gravitationally-captured matter can  be sufficiently large to form a disk before reaching the NS magnetosphere. 

Some early works \cite[e.g.][]{ghosh78,ghosh79} assumed that the NS magnetic field can penetrate the disk over a large range of radii, forming a relatively wide transition zone. It was later shown that this scenario requires an unphysically high magnetic diffusivity in the accretion disk \cite[see, e.g.,][]{wang95,lesur09}. Calculations assuming a more realistic magnetic diffusivity indicate that the NS magnetic field can penetrate the disk only over a quite small range of radii \citep{campbell92,shu94,lovelace95}, comparable to the thermal disk scale height $h\sim c_s/\omega_{\rm K}$, where $c_s$ is the sound speed  in the disk and $\omega_{\rm K}$ is the Keplerian frequency. In these models the inner part of the flow is frozen into the closed field lines magnetosphere, and the outer part of the flow contains open field lines leading to the possibility of magnetically driven outflows (see Fig.\ref{schemes}).

The thickness of this transition region influences the structure of the accretion column onto the NS.
If the area of the footprint of the accretion flow is small, the release of energy brought by the flow can be locally super-Eddington and can create a radiative shock which stops the infalling plasma. 
Despite significant theoretical efforts \cite[e.g.][]{basko76,ghosh78,ghosh79,campbell92,shu94,lovelace95} the width of the transition region remains uncertain, and until recently could not be estimated observationally. 
A method to estimate the thickness of the transition region  based on measurements of the cooling time of matter in the accretion column has been proposed for magnetic white dwarfs \citep{semena12,semena14}. The inferred plasma penetration depth at the boundary of the magnetosphere is in general agreement with earlier theoretical   \citep{shu94,lovelace95}. Extending these results to accreting magnetic NSs indicates that the footprint of their accretion columns might occupy an area as small as $<10^{-6}$ of the total NS surface. 

\subsection{Spin-period variations}

Matter that moves toward the NS carries angular momentum, which can be  added to that of the NS. If the accretion occurs via a Keplerian  disk, the spin-up torque acting on the NS can be estimated as $K_{\rm su}\approx\dot{M}\sqrt{GMr_{\rm m}}$ \citep{pringle72}.
On the other hand, the rotating magnetosphere can lose angular momentum via different mechanisms. For example, it can  transfer  angular momentum via large scale magnetic field threading the external regions of the accretion disk, which have a slower rotation \cite[e.g.][]{ghosh79,lovelace95,wang95}, or it can lose angular momentum by some matter outflow \cite[e.g.][]{illarionov90}. The spin-down torque $K_{\rm sd}$ acting on the NS depends on how the accretion in the binary occurs (e.g. via Keplerian disk or via quasi-spherical flow) and is still a matter of debate \cite[see e.g.][]{davidson73,kundt76,lipunov81,bk91,wang95,li96,kluzniak07,shakura12}. Leaving aside these complications, we can write the evolution of the spin period of NS accreting from a disk in  a generalized form:

\begin{equation}
\label{spin_evolution}
I {{d\omega}\over{dt}}=k_{\rm m}\dot{M}\sqrt{GMr_{\rm m}}-K_{\rm sd}
\end{equation}

\noindent
where $k_{\rm m}$ is a numerical constant of the order of unity.

If the spin-up torque is significantly higher than the spin-down torque (e.g. during episodes of high mass accretion rate), the dependence of the spin-up rate on the source luminosity/mass accretion rate can be expressed as follows:

\begin{equation}
\dot{\nu}\sim{\sqrt{GMr_{\rm m}}\over{2\pi 0.4 MR^2}} \dot{M}\sim {G^{3/7}\mu^{2/7}\dot{M}^{6/7}\over{2\pi 0.4 M^{4/7} R^2}}
\label{spinup_formula}
\end{equation}

\begin{equation}
\dot{\nu}\sim 3.5\times10^{-13} \mu_{30}^{2/7} \left({\dot{M}\over{10^{16} ~~\textrm{g s}^{-1}}}\right)^{6/7} ~ \rm{s^{-2}}
\end{equation}

\begin{equation}
\dot{\nu}\sim 2.5\times10^{-13} \mu_{30}^{2/7} \left({L_{\rm X}\over{10^{36} ~~\textrm{erg s}^{-1}}}\right)^{6/7} ~ \rm{s^{-2}}
\end{equation}

\noindent
The time scale $t_{\rm su}={\nu/{\dot{\nu}}}$ for spinning-up the accreting NS is  

\begin{equation}
t_{\rm su}\sim 3\times10^{5} ~\nu^{4/3} \left({10^{37}~\rm{erg~s}^{-1}\over{L_{\rm X}}}\right) \left({r_{\rm co}\over{r_{\rm m}}}\right)^{1/2}\rm{yr}
\end{equation}
 
\noindent
For typical mass accretion rates occurring in  high-luminosity sources  ($10^{37}-10^{38}$ erg s$^{-1}$), this time scale is short and spin-up should be clearly observable. Indeed, the spin-up of   Cen X-3 was already revealed since its first observations (see \citealt{bildsten97} for a compilation of spin-period time histories of  several accretion-powered pulsars). 

\begin{figure}[htb]
\includegraphics[width=\columnwidth,bb=234 162 694 524,clip]{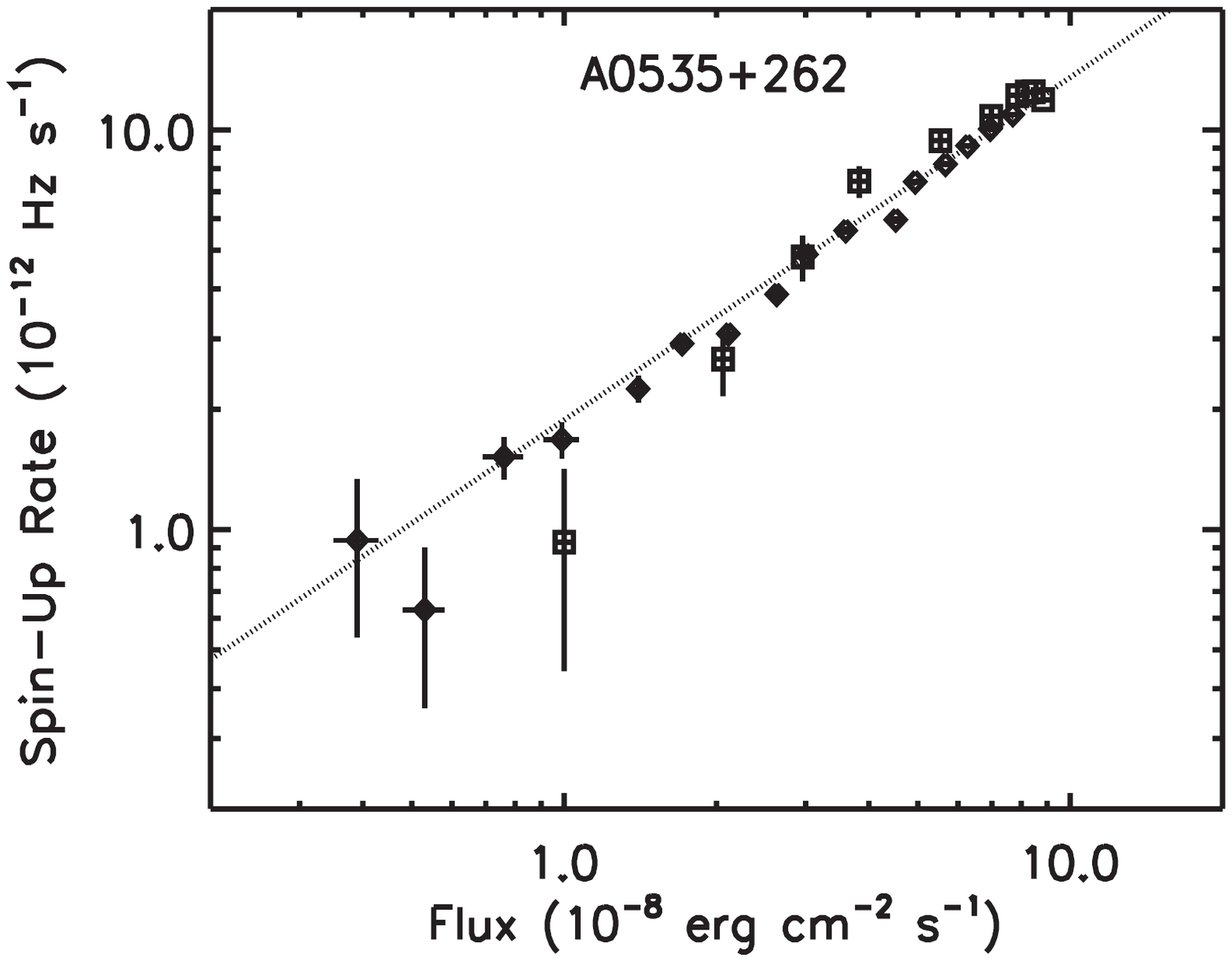}
\caption{Dependence of NS spin-up rate on X-ray flux (i.e. mass accretion rate). The dotted line indicates the prediction of the simple model of interaction of a dipole magnetosphere with a Keplerian disk (from \citealt{bildsten97}).}
\label{a0535_spinup}
\end{figure}

To illustrate the dependence of spin-up rate on  X-ray luminosity outlined in eq. (\ref{spinup_formula}),  let us consider the case of A0535+262, one of the brightest HMXB transients. In this system the NS  accretes from its Be-type companion via a thin disk. The mass accretion is not continuous, but it occurs via rare violent outbursts, during which the accretion rate changes by more than three orders of magnitude, resulting in  luminosities from $(0.5-1)\times10^{34}$ to $\sim10^{38}$ erg s$^{-1}$. The dependence of its spin-up rate on the  X-ray luminosity  gives us a direct probe of the model of disk-magnetosphere interaction and of the properties of the magnetosphere. When A0535+262 is at high luminosity, the accretion disk squeezes the magnetosphere to sizes much smaller than the corotation radius, thus reducing the flow-magnetosphere interaction to the simple scenario outlined above.
Fig.\ref{a0535_spinup}  shows how the observed spin-up rate depends on the X-ray flux, as observed during the bright phase of an outburst which occurred in 1996 \citep{bildsten97}. The dotted line indicates the relation $\dot{\nu}\propto \dot{M}^{6/7}$, deduced from the simple model described above. The observed points closely follow the model prediction and yield a NS surface magnetic field $B\sim 10^{12}$ G, which roughly agrees with the value derived from studies of cyclotron absorption lines.

A complete description  of  the spin-up/spin-down behaviour of accreting NSs is much more complicated than the simple picture outlined above, owing to  the interaction between  the accretion flow and the magnetosphere, which is particularly complex when the NS is rotating close to the equilibrium period, i.e. when $r_{\rm m}\approx r_{\rm co}$, and in the propeller stage, when $r_{\rm m}> r_{\rm co}$. These difficulties  are illustrated, e.g., by the case of KS 1947+300, for which the magnetic field derived from the $\dot{\nu}-L_{\rm X}$ correlation \citep{tsygankov05} is more than one order of magnitude higher than that indicated by the cyclotron line detected in this source.

\begin{figure}[htb]
\includegraphics[width=\columnwidth,bb=76 22 780 555,clip]{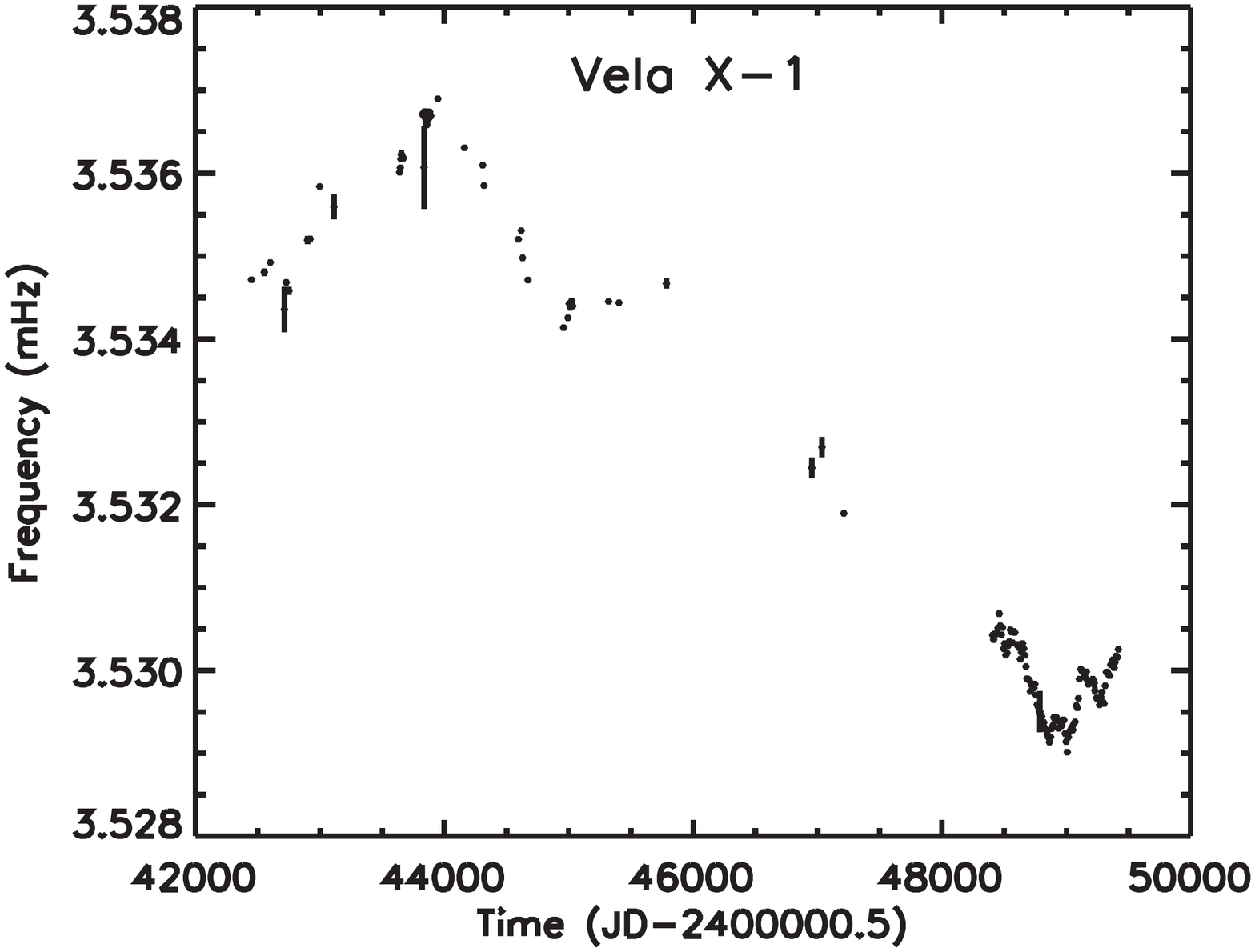}
\includegraphics[width=\columnwidth]{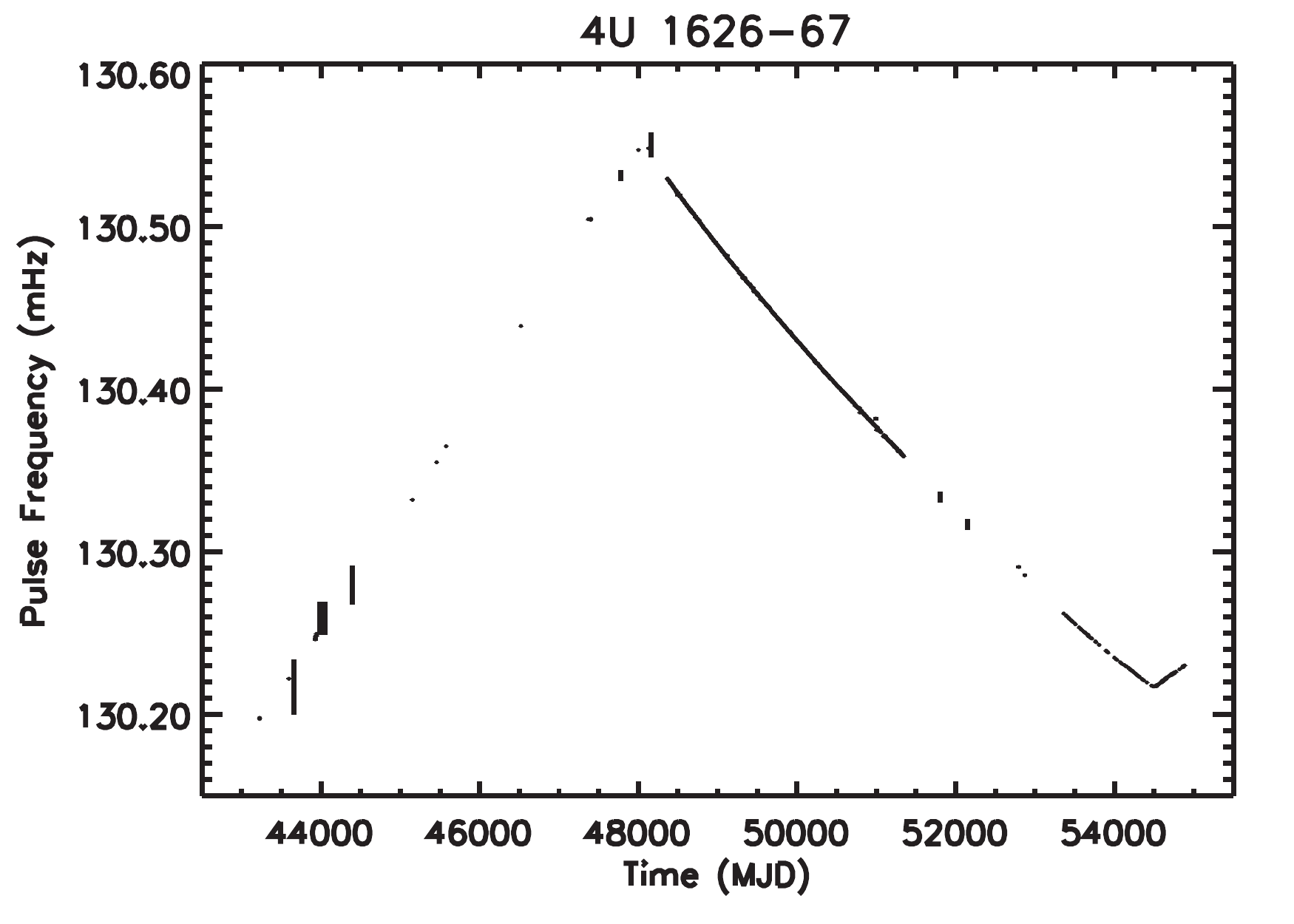}
\caption{Examples of spin-up/spin-down transitions of the wind accreting magnetic NS in Vela X-1 (top panel, from \citealt{bildsten97}) and the disk accreting magnetic NS in 4U 1626--67 (bottom panel, from \citealt{camero2010}).}
\label{1626_spinup}
\end{figure}

The existence of counteracting torques  (eq. (\ref{spin_evolution})) implies that: 1) there should be time periods of torque reversal, and 2) NSs should have some equilibrium spin periods.  
Indeed, observations do demonstrate time periods with zero time derivative of the NS spin value in some disk-accreting systems \cite[e.g.][]{parmar89,wilson02,baykal02}. 
Alternating periods of spin-up and spin-down are common in wind-accreting systems, where the angular momentum of the gravitationally-captured matter is subject to large fluctuations of both signs (see, e.g., the spin history of the wind accreting NS in Vela X-1 in the upper panel of Fig.\ref{1626_spinup}), but they have also been observed in NSs which are known to accrete through a disk (an example is shown in the bottom panel of Fig.\ref{1626_spinup}). A comprehensive review of observational facts about spin-up and spin-down histories can be found in, e.g., \cite{bildsten97}.

In these cases the observed luminosity/mass accretion rate values were used to make an estimate of the NS magnetic field, adopting some prescription of the spin-down torque $K_{\rm sd}$. Application of this approach to  A0535+262 provides a NS magnetic field value in agreement with that known from position of cyclotron absorption line \citep{doroshenko14}.
The problem of torque reversals in wind-accreting (without Keplerian accretion disk) binaries has been recently re-examined in the context of the quasi-spherical subsonic accretion regime by \citep{shakura12}. Application of this model to Vela X-1 and GX 301--2 allowed these authors to estimate magnetic fields  consistent with those derived from the cyclotron lines.

\section{Aperiodic X-ray flux variations as a tool to measure NS magnetospheres}
\label{sec:aperiodic}

The availability of high-quality data on accreting X-ray binaries has made it possible to investigate the NS magnetospheres using new methods based on the study of their aperiodic  X-ray variability.
Aperiodic variability in the X-ray flux of accreting sources was discovered at the dawn of X-ray astronomy \cite[e.g.,][]{oda71} and since then it has been extensively studied, mostly in the frequency domain, through the use of Fourier analysis techniques \citep{vanderklis89}.  The noise power spectrum in accreting sources has been shown to follow a power law in broad range of frequencies.  

\begin{figure}[htb]
\includegraphics[width=\columnwidth,bb=90 1 656 570,clip]{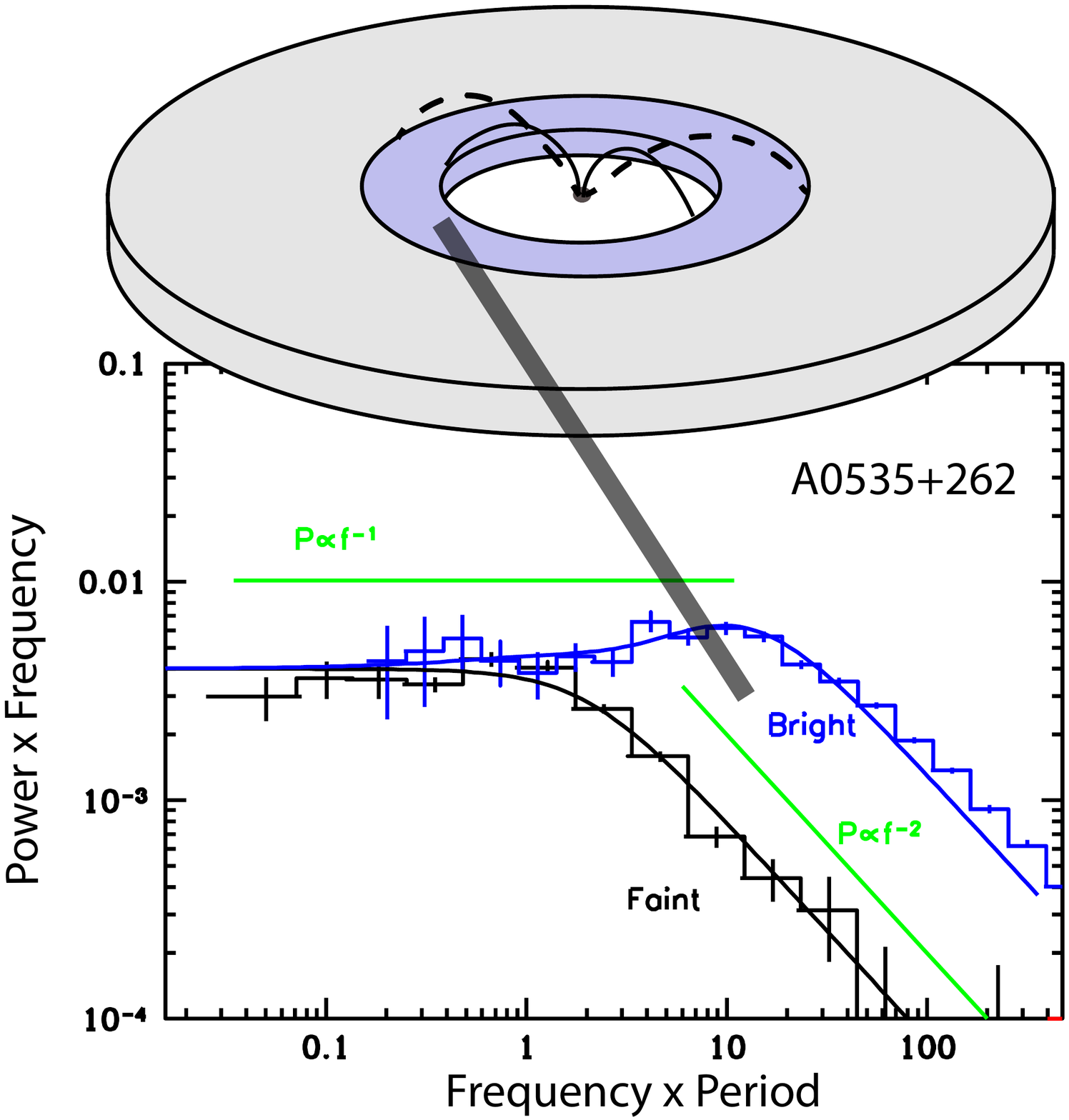}
\caption{Upper part: scheme of the accretion disk around a NS magnetosphere at two levels of the mass accretion rate. At high accretion rate the inner radius of the disk shrinks and an additional part of the flow with the fastest noise appears. Lower part: power spectra (pulse variations removed) of the accreting X-ray pulsar A0535+262 at different X-ray luminosities. The thick line relates the inner part of the disk with the fastest variability, not present during  low accretion rate periods, when the magnetosphere pushes the disk further away from the NS.}
\label{a0535_twostates}
\end{figure}

It is now  widely accepted that  this variability originates from the modulation of the instantaneous values of the mass accretion rate at different distances from the central object and their subsequent multiplicative superposition (so called model of propagating fluctuations; \citealt{lyubarsky97,churazov01,arevalo06,revnivtsev08}). In the framework of this model, the  varying luminosity originates in the central parts of the accretion flow (i.e. the accretion column in the case of X-ray pulsars),  but the modulations of the mass accretion rate are inserted into the flow at different (including large) distances from the central object, as a result of the stochastic nature of the viscosity in accretion disks (see, e.g., \citealt{balbus91,armitage03,hirose06}). 

The fastest variations, originating  closer to the compact object,  modulate the accretion rate in the disk, which is incoming to these regions from the outer disk regions. This model predicts that, if the accretion disk is truncated at some distance from the compact object, the broad band power spectrum of its luminosity variations should show a break (or steepening) resulting from the lack of variability at the highest frequencies (see Fig.\ref{a0535_twostates}). The break frequency, $f_{\rm b}$, should correspond to the time scale at which the inner part of the accretion flow modulates the mass accretion rate. As the truncation radius moves inward,  $f_{\rm b}$ should increase. Observational support of this prediction was demonstrated by \cite{revnivtsev09}.

\begin{figure}[htb]
\includegraphics[width=\columnwidth,bb=160 60 700 590,clip]{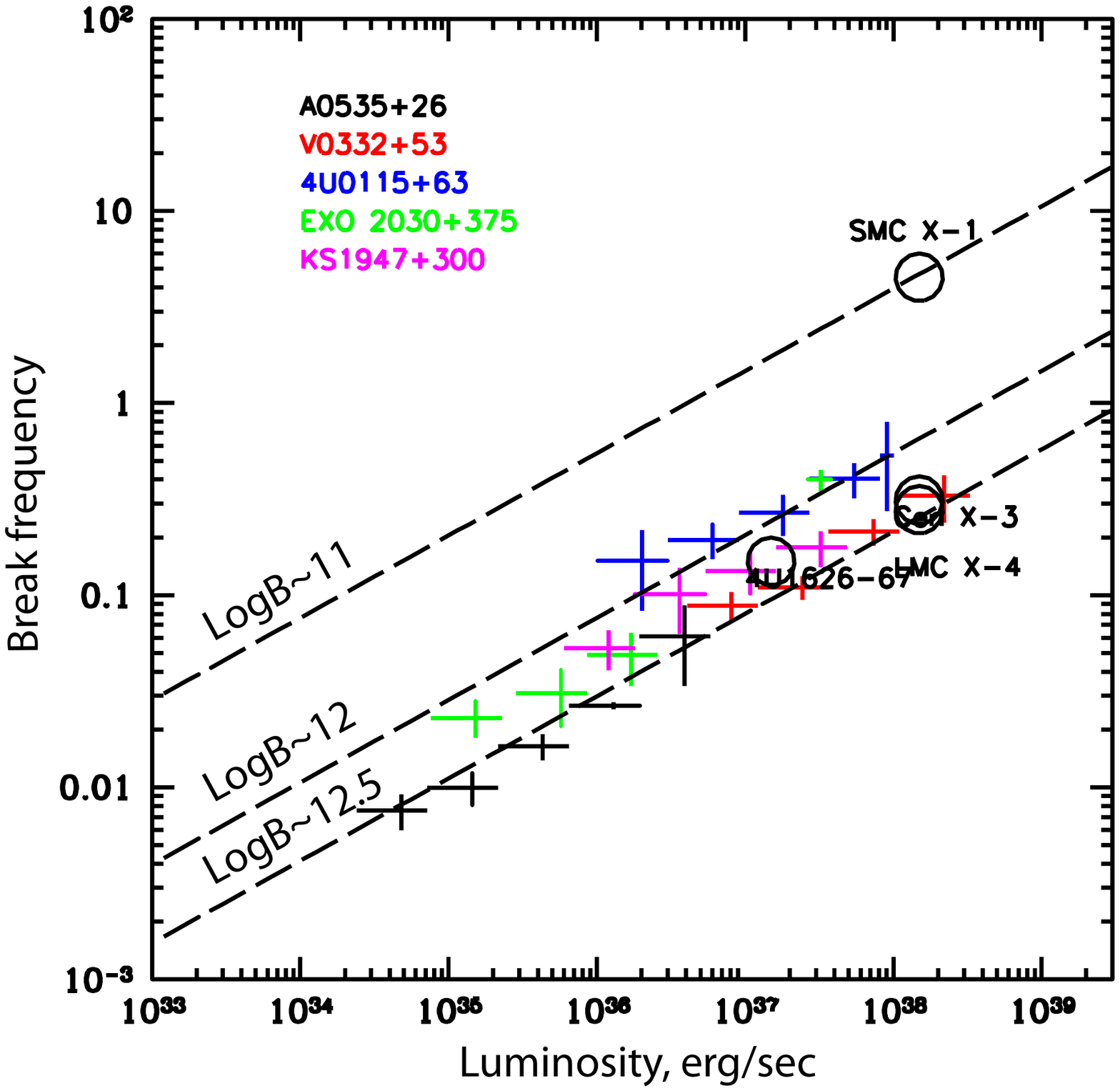}
\caption{Luminosity dependence of the break frequency in power spectra of disk accreting X-ray pulsars. Dashed lines denote approximate positions of breaks predicted by a simple model of Keplerian disk-dipole magnetosphere interaction with different strengths of NS surface magnetic field. The exact position of these lines should be treated  with caution due to uncertainties of NS parameters and properties of the magnetosphere.}
\label{fbreak_lx}
\end{figure}

If the characteristic frequency $f_0$ of the noise at the magnetospheric boundary $r_{\rm m}$ is proportional to the frequency of the Keplerian rotation $\nu_{\rm K}$ of matter at the inner edge of the accretion disk $r_{\rm in}\approx r_{\rm m}$, we can relate the observed break frequency to the instantaneous value of the mass accretion rate $\dot{M}$:
\begin{equation}
2\pi \nu_{\rm K} = (GM)^{1/2} r_{\rm m}^{-3/2}
\end{equation}

\noindent
and the break frequency will follow the dependence:
\begin{equation}
f_{\rm b}\propto f_0\propto\nu_{\rm K}(r_{\rm in})\propto (GM)^{5/7} \mu^{-6/7} \dot{M}^{3/7}
\label{fb}
\end{equation}

\noindent
This relation was observed in transiently accreting pulsars \cite[see e.g.][and Fig.\ref{fbreak_lx}]{revnivtsev09,tsygankov12,doroshenko14}.

In addition to the broad band noise,  accreting X-ray pulsars sometimes show quasi-periodic oscillations (QPO) of their fluxes \cite[see e.g.][]{shinoda90,finger96,cameroarranz12}.
It was found that the QPO frequency varies during the outbursts of the transient A0535+262 and its value strongly depends on the   X-ray flux. It is remarkable that the dependence of the QPO frequency on the accretion rate is very similar to that of the frequency of the break in the power spectrum of the source flux variability, which hints on their common origin. It was proposed that the QPO might originate at the inner boundary of the accretion disk, truncated by the NS magnetosphere at the beat frequency between the Keplerian frequency at disk inner edge and the neutron star rotation frequency \citep{finger96}.

\section{Magnetic propeller effect}
\label{sec:propeller}

Due to the increase of magnetospheric radius with the decrease of the mass accretion rate (eq. (\ref{rm})),   at some stage of transiently accreting NSs (e.g. in  Be HMXBs  during outbursts or in LMXB transients) the magnetospheric radius can reach the corotation radius and stop the direct accretion regime. This is the so called ``propeller effect''.  The limiting mass accretion rate $\dot{M}$, and the corresponding luminosity,  can be estimated by the  condition  $r_{\rm m}\sim r_{\rm co}$:

\begin{equation}
\dot{M}\sim \xi^{7/2} {(2\pi)^{7/3} \mu^2 \over{G^{5/3}M^{5/3}P^{7/3}}}
\end{equation}

\begin{equation}
L_{\rm x}\sim \xi^{7/2} 2.3\times10^{33} \mu^{2}_{30}  \left({100~\rm{s}\over{P}}\right)^{7/3}~~~\textrm{erg} ~\textrm{s}^{-1}
\end{equation}

\noindent
Below we consider several consequences of the propeller effect which can be probed with observations.

\begin{figure}[htb]
\includegraphics[width=6.5cm,angle=-90]{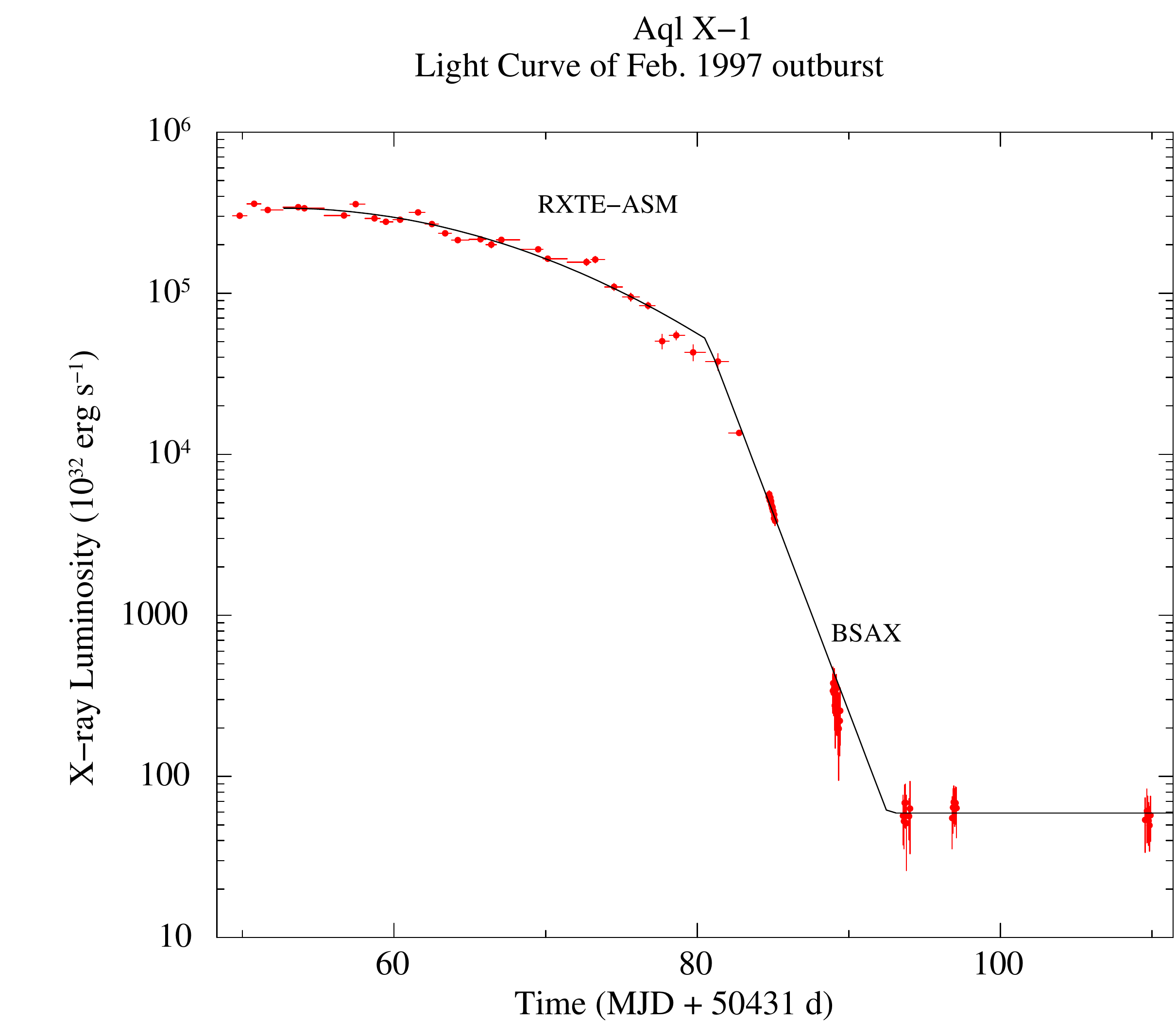}
\caption{Light curve of the 1997 outburst of the LMXB transient Aql X-1. After a relatively smooth decline, an abrupt drop of X-ray luminosity can be seen around $L_{\rm x}\sim5\times10^{36}$ erg s$^{-1}$. This behavior can be explained by the the propeller effect of a rotating magnetosphere  with  $B\sim(1-4)\times10^{8}$ G  at the NS surface (from \citealt{campana14}).}
\label{propeller}
\end{figure}

\subsection{Luminosity drops in HMXBs and LMXBs}

When the magnetospheric radius starts to exceed the corotation radius, one should expect to see an abrupt drop of the source luminosity or a switch to unsteady accretion with large luminosity swings. Indeed, at this stage the matter gains angular momentum from the rotating magnetosphere and can not move directly toward the NS. 
Relatively small variations of the incoming mass rate can lead to large variations of the X-ray luminosity.

Such drops were indeed observed during the decreasing phases  of the oubursts of several Be binaries, like,  e.g. V0332+53 \citep{stella86}, 4U0115+63 \citep{campana01}, Vela X-1 \citep{doroshenko11}, 4U1907+09 \citep{doroshenko12}. This effect is not easy to catch because it occurs in a short time interval, when the source is crossing the above luminosity limit. An additional complication comes from the fact that when XRBs reach low luminosity levels ($L_{\rm x}< 10^{33}-10^{34}$ erg s$^{-1}$), the  contribution provided by X-ray emission of the companion star may become non-negligible. 

It is possible that the pure propeller effect, causing the complete ejection of gravitationally captured matter, is unlikely to be realized unless the compact magnetized rotator is really fast (as in the case of the rapidly rotating magnetized white dwarf in the binary AE Aqr, see, e.g., \citealt{wynn97}). In the case of slow rotators some residual accretion might still be possible from a quasi-spherical reservoir above the magnetosphere (see e.g. \citealt{shakura12}) or from so-called ``dead-disks" \citep{illarionov75,dangelo10,dangelo12}.

If the mass supply rate at the magnetospheric boundary persists for some time, one might expect large amplitude quasi-cyclic variations of X-ray luminosity, due to intermittent matter penetration \citep{baan79,dangelo10,dangelo12}. Such variations were observed in several cases: A0535+262 \citep{caballero08,postnov08}, EXO 2030+375 \citep{klochkov11}.

The influence of the propeller effect has  also been observed in transiently accreting LMXBs (Fig.\ref{propeller}), where abrupt drops of X-ray luminosity were seen in the deacying parts of the outbursts  \cite[see, e.g.,][]{cui97,gilfanov98,asai13,campana14},  leading to  estimated magnetic fields of a few $10^{8}$ G.   Note, that in the case of fast pulsars in LMXBs the inner edge of the disk is much closer to the NS surface than in HMXBs. Thus the luminosity variations associated to the propeller effect are smaller and more difficult to detect.  Also considering that the determination of $r_{\rm m}$ is subject to larger uncertainties in these systems, the above estimates should be taken with some caution \cite[see, e.g., discussion in][]{patruno12}. 

\subsection{Influence of the propeller effect on XRBs populations}

The presence of rotating NS magnetospheres should prevent persistent accretion at low rates (see, e.g., discussion of this point in \citealt{lipunov82,stella86,shtykovskiy05}; numerical simulation predict similar effect in LMXB populations, see, e.g., \citealt{kuranov14}). 
In principle, this prediction could be tested with statistical analysis of well defined samples of XRBs, which should reveal a lack of objects in  some regions of parameters (e.g. luminosity, spin-period, magnetic field, orbital period). However, such studies are complicated by observational biases and by large uncertainties on many parameters. 
 
Attempts to obtain a well defined sample of X-ray binaries with the help of sky surveys can be traced back to the first X-ray sky surveys \cite[see e.g.][]{matilsky73}. However, it took a long time to measure the relevant physical parameters of these binaries (orbital period, distance, etc...), which are needed for quantitative comparisons with the predictions of population synthesis models.

The simplest comparison between observations and population synthesis
can be done for the distribution of  X-ray luminosities. This is more difficult do for the sources in our Galaxy, due to the uncertainties on their distances, but it is much easier for sources in the   Magellanic Clouds.  \cite{shtykovskiy05} have shown that the flattening below $L_{\rm x}<10^{35}$ erg s$^{-1}$ of the luminosity function of LMC  X-ray sources  might indicate an influence of the propeller effect. More solid confirmation of this effect is required on much higher statistics of sources at such low luminosities.

Population studies can also be used to test some of the models proposed to explain the SFXTs invoking the role of the rotating magnetosphere in the inhibition of accretion \cite[e.g.][]{grebenev07,bozzo08}. In these models the matter gradually accumulates above the magnetospheric boundary and occasionally finds its way to the NS surface, giving rise to the bright flares which characterize these sources.

A simple picture of stellar wind accretion predicts the existence of an ``allowed" region in  the $P_{\rm orb}-L_{\rm x}$ parameter space,  determined by the minimum orbital period of a NS with a massive companion not filling its Roche lobe, and the minimum mass accretion rate which can be supplied by such a massive star \citep{bhadkamkar12,lutovinov13}.
This is based on the fact that the accretion rate onto a  NSs at distance $a$  from a companion of mass $M_2$ depends almost only on the mass loss rate of the secondary $\dot{M}_{2}$: $L_{\rm x}\propto a^{-2} f(M_2)$. For OB stars with wind velocities $v_{\rm w}\sim 1000$ km s$^{-1}$,  the mass loss rate $\dot{M}_{2}\propto M_2^{\alpha}$,  with $\alpha\sim$2.76. Therefore, for any fixed separation $a$ (i.e. approximately fixed $P_{\rm orb}$),  the NSs with the smallest mass companions should have the lowest X-ray luminosities. 
This lower boundary on $L_{\rm x}$ for persistently accreting sources should have a functional form $L_{\rm x}\propto a^{-2}\propto P_{\rm orb}^{-4/3}$. 
If an X-ray luminosity below this limit is observed, some additional mechanism must be invoked to reduce the NS mass accretion rate.

\begin{figure}[htb]
\includegraphics[width=\columnwidth,bb=4 34 555 586,clip]{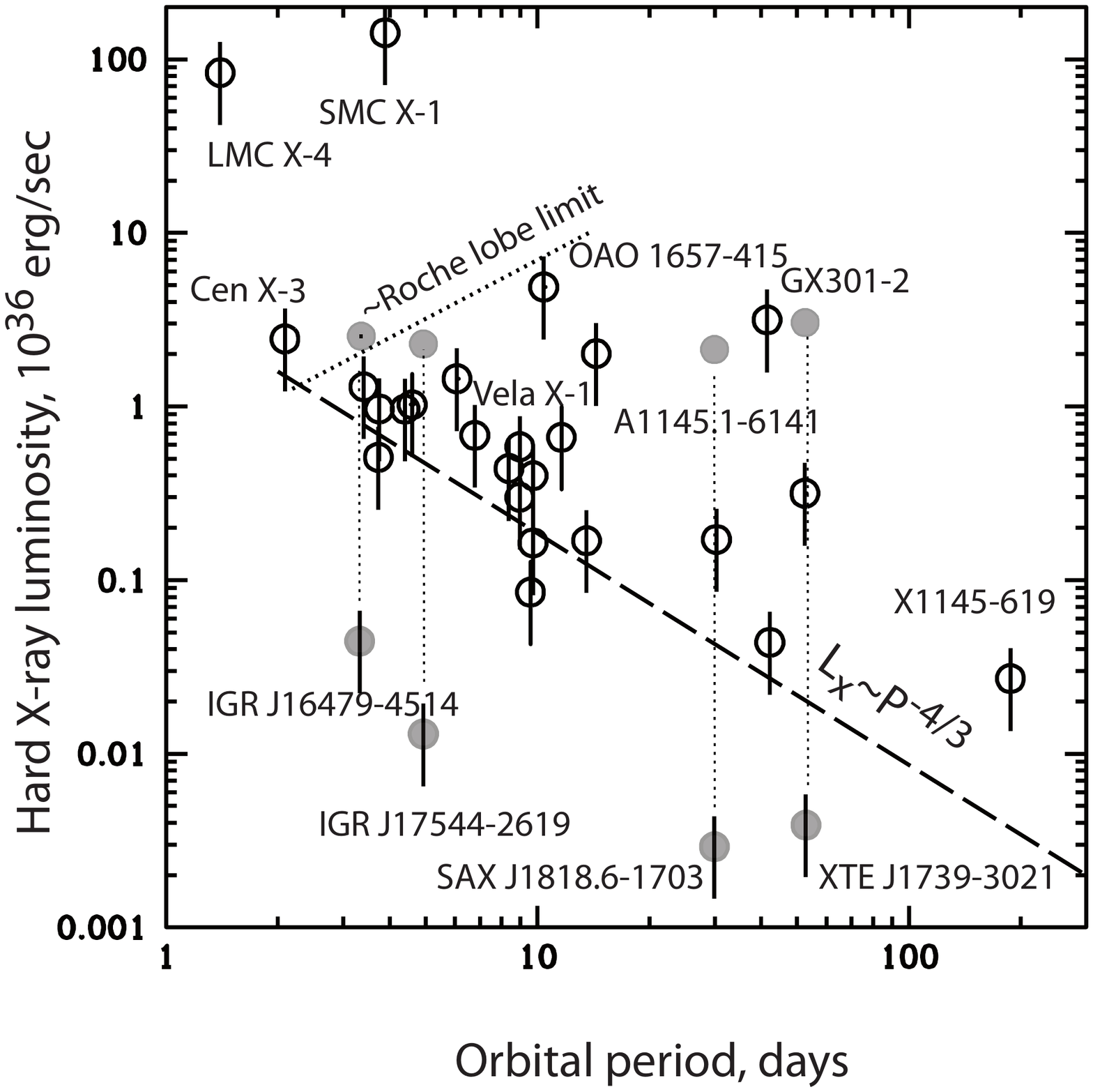}
\caption{Hard X-ray luminosity versus orbital period of persistent HMXBs of our Galaxy.  Sources below the dotted line accrete from stellar wind, sources above this line accrete via Roche lobe overflow. The dashed line denotes the approximate lower limit of {\sl persistent} X-ray luminosity which can be provided by accretion from minimally outflowing massive stars (i.e. from young stars with masses around $M_2\sim 7-8 M_\odot$). Filled circles connected with vertical dotted lines denote positions of supergiant fast X-ray transients at their average  (lower circle) and maximum  (upper circle) luminosity state. Adapted from \citet{lutovinov13}.}
\label{lx_porb}
\end{figure}

This prediction  can be tested using the currently best available  flux-limited sample of accreting NSs in  HMXBs presented in \cite{lutovinov13}. This survey is complete over the whole Galaxy down to luminosities $(0.5-1)\times 10^{35}$ \lum\ in the energy range 17-60 keV.
As shown in Fig. \ref{lx_porb} the persistent NSs with OB giant companions in our Galaxy populate the ``allowed" area in the  $P_{\rm orb}-L_{\rm x}$ diagram. On the other hand,  a few sources  belonging to the class of supergiant fast X-ray transients lie  below the   $L_{\rm x}$ lower boundary. 
According to above mentioned arguments, this suggests that  that  mass accretion onto the NS surface is inhibited (or strongly reduced) as a result of magnetospheric/centrifugal inhibition of accretion.

\section{Conclusions}

Since the time of their discovery in the late 60's,  XRBs have been used as natural laboratories for studies of matter in conditions of extreme density and magnetic fields. A substantial fraction of them  contains accreting  NSs  endowed with magnetic fields sufficiently strong to significantly affect their observed properties.  The presence of periodic X-ray pulsations provided immediate evidence for magnetic fields of the order $\sim10^{12}$ G in the HMXBs, similar to those deduced in a completely independent  way from  the spin-down rate of young radio pulsars.  At the same time, the rarity of pulsations in LMXBs pointed to much smaller magnetic fields for the NSs contained in these, generally older, systems.    

The detection of cyclotron resonance features gives the most direct way to estimate the magnetic field in XRBs. The wealth of  good observational data on XRB cyclotron lines is now posing challenges to the theory developed for the spectral formation in accreting magnetized NSs, despite the  high level of complexity now reached by these models. However, despite some difficulties in the detailed modelization, it is well established that the observed line energies correspond to magnetic fields in the range $\sim10^{12}-10^{13}$ G.

Other ways to estimate the magnetic field in XRBs are less direct than cyclotron line measurements (and often more model-dependent), but have the advantage that they can be applied over a larger range of magnetic field intensities.  The study of spin-up and/or spin-down rates in connection with the source luminosity gives the possibility to estimate the large scale magnetic field of the NS, because the exchange of angular momentum between the NS and the surrounding matter is strongly mediated by the magnetic field.  The results obtained by these analysis are rather well established in the case of disk accretion during  the bright outbursts of transient XRBs, spanning a large range of accretion rate values. In the case of (quasi-)spherical matter flows  and/or  more complex regimes  (e.g. propeller, intermittent accretion, dead accretion disks, etc...), the results are subject to larger uncertainties.   Other successful techniques  to estimates NS magnetic field strengths rely on the observation of sudden luminosity drops due to the onset of the propeller effect  and on the study of rapid aperiodic variability.  The first method proved particularly valuable in the case of LMXB transients, where fields of the order of a few $10^{8}$ G   were derived.

In general, there is a reasonably good agreement between the  magnetic fields estimated with alternative methods and those obtained from  cyclotron-line measurements. However, detailed comparisons have been possible only in a limited number of cases and they often rely on some poorly known quantities (e.g., the mass accretion rate,  the geometry of the magnetosphere, the magnetization of the accreting plasma). These uncertainties, coupled to some poorly justified assumptions, can explain some discrepant magnetic field estimates reported in the literature.  It must also be remembered that the distribution of field values derived from cyclotron lines is affected by the current instrument capabilities, which make it difficult to detect narrow spectral features at energies above a few tens of keV and below $\sim$1 keV. Improvements in this field will be   obtained by future missions with a better energy resolution in the soft X-ray range and higher sensitivity at hard X-rays.

 \begin{acknowledgements}
We thank all the staff of the International Space Science  Institute and the organizers of the stimulating  Workshop ``The Strongest Magnetic Fields in the Universe''. 
MR acknowledges the support by grant RNF 14-12-01287.  SM has been supported through financial contribution from the agreement ASI-INAF I/037/12/0.
\end{acknowledgements}

\bibliographystyle{aa}      
\bibliography{bibreview}

\end{document}